\documentclass[a4paper,12pt]{article}
\usepackage{graphicx}

\def\beq{\begin{equation}}
\def\eeq{\end{equation}}
\def\bea{\begin{eqnarray}}
\def\eea{\end{eqnarray}}

\def\roughly#1{\mathrel{\raise.3ex\hbox
{$#1$\kern-.75em\lower1ex\hbox{$\sim$}}}}

\def\c{{\cal C}}

\def\b{{\cal B}}
\def\acp{{\cal A}_{\rm CP}}

\def\l{\lambda}

\def\cbrg{{B^\pm \rightarrow \rho^\pm \gamma}}

\newcommand{\abseps}{\vert\epsilon\vert}
\newcommand{\delmd}{\Delta M_d}
\newcommand{\delms}{\Delta M_s}

\def\journal#1#2#3#4{{\it #1} {\bf #2} (#3) #4}
\def\epj{Euro. Phys. Jour.}
\def\prl{Phys. Rev. Lett.}
\def\pl{Phys. Lett.}
\def\np{Nucl. Phys.}

\def\zp{Z. Phys.}
\def\pr{Phys. Rev.}

\begin{document}

\begin{flushright}
DESY 00-088 \\
UdeM-GPP-TH-00-75\\
June 2000\\
\end{flushright}

\begin{center}

  {\Large \bf 
\centerline{Supersymmetric Effects on Isospin Symmetry} 
\centerline{Breaking and Direct CP Violation in $B \to \rho \gamma$}} 
\vspace*{1.5cm}
{\large A.~Ali, ~L.T. Handoko} \vskip0.2cm
Deutsches Elektronen Synchrotron DESY, Hamburg \\
\vspace*{0.3cm} \centerline{ and} \vspace*{0.3cm} {\large D.~London}
\vskip0.2cm Laboratoire Ren\'e J.-A. L\'evesque, Universit\'e de
Montr\'eal, \\
C.P. 6128, succ.\ centre-ville, Montr\'eal, QC, Canada H3C 3J7 \\
\vskip0.5cm 
{\Large Abstract\\} 
\vskip3truemm 

\parbox[t]{\textwidth} {We argue that one can search for physics
  beyond the standard model through measurements of the
  isospin-violating quantity $\Delta^{-0} \equiv \Gamma(B^- \to \rho^-
  \gamma)/2\Gamma(B^0 \to \rho^0 \gamma)-1$, its charge conjugate
  $\Delta^{+0}$, and direct CP violation in the partial decay rates of
  $B^\pm \to \rho^\pm \gamma$. We illustrate this by working out
  theoretical profiles of the charge-conjugate averaged ratio $\Delta
  \equiv {1 \over 2}(\Delta^{+0} +\Delta^{-0})$ and the CP asymmetry
  ${\cal A}_{CP}(B^\pm \to \rho^\pm \gamma)$ in the standard model and
  in some variants of the minimal supersymmetric standard model. We
  find that chargino contributions in the large $\tan \beta$ region
  may modify the magnitudes and flip the signs of $\Delta$ and ${\cal
    A}_{CP}(B^\pm \to \rho^\pm \gamma)$ compared to their
  standard-model values, providing an unmistakeable signature of
  supersymmetry.}

\end{center}
\thispagestyle{empty}
\newpage  
\setcounter{page}{1}
\textheight 23.0 true cm

Measurements of the radiative decays $B \to K^* \gamma$
\cite{cleobkstar} and $B \to X_s \gamma$ \cite{cleobsg} have triggered
a large number of theoretical studies whose aim is to provide
precision tests of the flavor sector in the standard model (SM), and
to search for possible hints of new physics, particularly
supersymmetry \cite{greub99}. The related Cabibbo-suppressed decays $B
\to \rho \gamma$, $B \to \omega \gamma$ and $B \to X_d \gamma$, for
which experiments have so far provided only upper bounds
\cite{cleobgrho}, but which surely will be measured at $B$-factories,
have also been studied at great length. Within the SM, these latter
decays are particularly interesting because they potentially allow us
to determine the Cabibbo-Kobayashi-Maskawa (CKM) matrix element
$V_{td}$, or, more generally, the quark mixing parameters $\bar\rho$
and $\bar\eta$ of the Wolfenstein parametrization of the CKM matrix
\cite{wolf}. While the inclusive decay is theoretically more robust
\cite{aag97}, it is experimentally very challenging. In view of this,
considerable effort has gone into consolidating the theoretical
profile of the exclusive decays $B \to V \gamma$ ($V=K^*,\rho,\omega)$
in the SM \cite{abs94,ksw95,ab95,dgp97,vmd,krsw97,gp00}.

In this letter, we argue that the interference of the short-distance
(SD) penguin amplitude and long-distance (LD) tree amplitude in
exclusive radiative $B$-decays, which is often considered as an {\it
  impediment} to a precise determination of the CKM parameters from
their branching ratios, may turn out to be {\it a boon in disguise} in
searching for new physics. To illustrate this point, we focus on the
decays $B^0(\overline{B^0}) \to \rho^0 \gamma$ and $B^\pm \to \rho^\pm
\gamma$, and consider the isospin-violating ratio defined as
\beq
\Delta^{-0} \equiv 
  {\Gamma(B^- \to \rho^- \gamma) \over 2 \Gamma(B^0 \to \rho^0 \gamma)} -1
~, 
\label{Deltadef}
\eeq
along with its charge conjugate $\Delta^{+0}$. (Since theoretical
estimates give $\tau(B^\pm) =\tau(B^0)$, to within a couple of
percent, and the present data support this conclusion \cite{pdg}, the
quantities $\Delta^{\pm 0}$ can be interpreted in terms of the
branching ratios.) Note that the ratios $\Delta^{\pm 0}$ deviate from
zero (their isospin limit) due to the SD-LD interference effects
mentioned above.

We compare the profiles of the charge-conjugate averaged ratio $\Delta
\equiv {1 \over 2}(\Delta^{+0} + \Delta^{-0})$ in the SM and in a
class of variants of the minimal supersymmetric standard model (MSSM)
in which all non-diagonal flavor transitions take place essentially
via the CKM quark mixing matrix. Although the SM and the MSSM yield
similar values of $\Delta$ in some regions of parameter space, in
other regions the MSSM may change this ratio significantly. This can
happen in two different ways. First, in some MSSM models a larger
value of the angle $\alpha$ in the unitarity triangle is preferred
\cite{al99}. Since $\Delta$ increases with $\alpha$ in the quadrant
$\pi/2 \leq \alpha \leq \pi$ \cite{ab95,dgp97}, the ratio $\Delta$ may
be enhanced in the MSSM. The second effect, which is particularly
striking, is that the sign of $\Delta$ can be flipped in MSSM models.
This can happen in that region of large $\tan\beta$ supersymmetric
parameter space in which the chargino-stop contributions are known to
flip the sign of the effective matrix elements of the electromagnetic
and chromomagnetic penguin operators \cite{agm94,cdgg,bghw,goto99}.

Finally, we also consider the direct CP asymmetry $\acp$ in the decays
$B^\pm \to \rho^\pm \gamma$. (The direct CP asymmetries in $B^0
(\overline{B^0}) \to \rho^0 \gamma$ and $B^0 (\overline{B^0}) \to
\omega \gamma$ are very similar to $\acp(B^\pm \to \rho^\pm \gamma)$,
though their time evolution will be modulated by
$B^0$--$\overline{B^0}$ mixing effects.) We find that, for MSSM's with
large $\tan\beta$, the sign of $\acp$ may turn out to be opposite that
of the SM.

Having summarized our main results above, we now present the
calculation. To compute the radiative weak transitions ($b \to d
\gamma)$, we use the effective Hamiltonian
\beq
{\cal H} = \frac{G_F}{\sqrt{2}}\left[\lambda_{u}^{(d)}(\c_1(\mu) {\cal
O}_1(\mu) + \c_2(\mu) {\cal O}_2(\mu)) -\lambda_{d}^{(t)} \c_7^{eff}(\mu)
{\cal O}_7(\mu) +...\right].
\eeq
Here, $\lambda_{q}^{(q^\prime)} = V_{qb}V_{qq^\prime}^*$ are the CKM
factors, and we have restricted ourselves to those contributions which
will be important in what follows. The operators ${\cal O}_1(\mu)$ and
${\cal O}_2(\mu)$ are the four-quark operators
\beq
{\cal O}_1 = (\bar{d}_\alpha \Gamma^\mu u_\beta)(\bar{u}_\beta
\Gamma_\mu b_\alpha) ~~,~~~~ 
{\cal O}_2 = (\bar{d}_\alpha \Gamma^\mu u_\alpha)(\bar{u}_\beta 
\Gamma_\mu b_\beta) ~, 
\eeq
where $\Gamma_\mu=\gamma_\mu(1-\gamma_5)$, $\alpha$ and $\beta$ are
the SU(3) color indices, and $\c_1$ and $\c_2$ are the corresponding
Wilson coefficients. ${\cal O}_7$ is the magnetic moment operator
\beq
{\cal O}_7 = \frac{e m_b}{8 \pi^2} \bar{d} \sigma^{\mu
\nu}(1-\gamma_5) F_{\mu \nu} b ~,
\eeq
where $F_{\mu \nu}$ is the electromagnetic field strength tensor.
We note that the coefficient $\c_7^{eff} (\mu)$ also includes the
effect of the four-quark operators ${\cal O}_5$ and ${\cal O}_6$, and
the operator matrix elements and their coefficients are calculated at
the $b$-quark mass scale $\mu=m_b$. For further details and
definitions, see Ref.~\cite{ab95}.

The decay amplitudes of interest can be written in the form:
\bea
{\cal M}(B^- \to \rho^- \gamma) &=& \lambda_{t}^{(d)}
a_{P}(1-\frac{\vert \lambda_{u}^{(d)}\vert}{\vert \lambda_{t}^{(d)}\vert}
R_{L}^{(-)} e^{i\alpha}) ~, \nonumber\\
{\cal M}(B^0 \to \rho^0 \gamma) &=& \lambda_{t}^{(d)}
a_{P}(1-\frac{\vert \lambda_{u}^{(d)}\vert}{\vert\lambda_{t}^{(d)}\vert}
R_{L}^{(0)} e^{i\alpha}) ~,
\eea
where isospin symmetry has been used in writing $a_P^{(-)}
=a_{P}^{(0)} \equiv a_P$ for the penguin amplitudes, and $\alpha$ is
one of the angles of the CKM unitarity triangle. The dynamical
quantities $R_{L}^{(-)}$ and $R_{L}^{(0)}$, which are in general
complex due to strong interactions, are the ratios of the reduced LD
and SD amplitudes in the decays $B^- \to \rho^- \gamma$ and $B^0 \to
\rho^0 \gamma$, respectively. In general, $a_P$, $R_{L}^{(-)}$ and
$R_{L}^{(0)}$ are all model-dependent.

Light-cone QCD sum rules, which take into account the dominant
$W^\pm$-annihilation and $W^\pm$-exchange contributions, typically
yield $R_L^{(-)} \simeq -0.3\pm 0.07$ and $R_L^{(0)} \simeq 0.03 \pm
0.01$ \cite{ksw95,ab95}. These estimates, which are obtained using the
factorization approximation, have been essentially confirmed by a
recent calculation in which non-factorizable corrections are proven to
vanish in the chiral limit to leading twist, in the heavy quark limit
\cite{gp00}.  In addition, long-distance contributions from other
topologies have been estimated systematically and found to be small
\cite{vmd,krsw97,gp00}.
Eventually, radiative decays $B^\pm \ell^\pm \nu \gamma$ can be used
to compute the leading ($W^\pm$-exchange) topologies in a
model-independent way \cite{gp00}. Of course, one still needs to know
$a_P$ to get the branching ratios.

The expression for the ratio of the branching ratios of interest can be
written as 
\beq
   \frac{\b(B^- \rightarrow \rho^- \gamma)}{
     \b(B^0 \rightarrow \rho^0 \gamma)} \simeq  
     \left| 1 + 
     \frac{4 \pi^2 m_{\rho^\pm}}{m_b} 
      \frac{\c_2 + {\c_1}/{N_c}}{\c_7^{eff}} r_u^{\rho^\pm} \,
\frac{\l_u^{(d)}}{\l_t^{(d)}} \right|^2 ~,
\eeq
where $r_{u}^{\rho^\pm}$ lumps together the dominant
($W$-annihilation) and possible sub-dominant LD contributions.
Borrowing the notation from Ref.~\cite{gp00},
\beq
     \epsilon_A {\rm e}^{i \phi_A} \equiv
        \frac{4 \pi^2 m_{\rho^\pm}}{m_b}
      \frac{\c_2 + {\c_1}/{N_c}}{\c_7^{eff}} r_u^{\rho^\pm} ~,
\label{phiAdef}
\eeq
and noting that $\lambda_{u}^{(d)}/\lambda_{t}^{(d)} = - \vert
\lambda_{u}^{(d)}/\lambda_{t}^{(d)} \vert e^{+i \alpha}$, which holds
in the SM and in the MSSM models being considered here, the isospin
breaking ratios [Eq.~(\ref{Deltadef})] can be expressed as
\beq
  \Delta^{\pm 0} = 2 \epsilon_A \left[
    \cos \phi_A F_1 \mp \sin \phi_A F_2 +
    \frac{1}{2} \epsilon_A (F_1^2 + F_2^2)
    \right] ~.
\eeq
Here, $F_{1,2}$ are (implicit) functions of the Wolfenstein parameters
$\bar\rho$ and $\bar\eta$:
\beq
F_1= - \left\vert {\l_u^{(d)} \over \l_t^{(d)}}\right\vert \cos\alpha 
~~,~~~~ 
F_2= - \left\vert {\l_u^{(d)} \over \l_t^{(d)}} \right\vert \sin\alpha ~,
\eeq
with $(F_1^2 + F_2^2) = \left\vert \l_u^{(d)} / \l_t^{(d)}
\right\vert^2$.

The charge-conjugated averaged ratio, defined as $\Delta \equiv
\frac{1}{2} \left[ \Delta^{-0} + \Delta^{+0} \right]$, has the
following leading-order (LO) expression:
\beq
 \Delta_{\rm LO} = 2 \epsilon_A \left[\cos \phi_A F_1 + \frac{1}{2}
\epsilon_A(F_1^2 + F_2^2)\right]
\simeq 2 \epsilon_A \left[F_1 +{1 \over 2}\epsilon_A(F_1^2 +F_2^2) \right] ~,
\label{eq:deltall} 
\eeq
where the near equality reflects that, in this approximation, the
strong interaction phase $\phi_A$ disappears in the chiral limit
\cite{gp00}.

In fact, one can go to next-to-leading-order (NLO) in the calculation
of the above quantities.  The NLO-corrected expression for the
branching ratios and $\Delta$ can be derived from the corresponding
calculations for the inclusive decay $B \to X_s \gamma$ \cite{ghw,cmm}
and $B \to X_d \gamma$ \cite{aag97}:
\bea
  \Gamma(\cbrg) & = & \frac{G_F^2 \alpha |\l_t^{(d)}|^2}{32 \pi^4} m_B^5 
  \left( 1 - \frac{m_\rho^2}{m_B^2} \right)^3 
  \left| T_1^{\rho} \right|^2 
    \left\{ 
    \left| \c_7^{(0)eff} + A_R^{(1)t} \right|^2  
    \right. \nonumber \\
    & & \left.
    + \left( F_1^2 + F_2^2 \right) \left( 
      \left| A_R^u + L_R^u \right|^2 \right)
    \right. \nonumber \\
    & & \left.
    + 2 F_1 \left[ \c_7^{(0)eff} 
      \left( A_R^u + L_R^u \right) + A_R^{(1)t} L_R^u \right] 
    \right. \nonumber \\
    & & \left.
    \mp 2 F_2 \left[\c_7^{(0)eff}  A_I^u 
     - A_I^{(1)t} L_R^u   \right] 
  \right\} \; .
\eea
Here $G_F$ is the Fermi coupling constant, $\alpha=\alpha(0)=1/137$,
$T_1^{\rho}$ is the $B \to \rho$ form factor involving the magnetic moment
operator ${\cal O}_7$, evaluated at $q^2=0$, and $L_R^u=\epsilon_A
\c_7^{(0)eff}$.  The quantities
$A_{R,I}^{(1)t}$ and $A_{R,I}^u$ represent the real and imaginary
parts of the explicit $O(\alpha_s)$ contributions to the matrix
elements evaluated at a scale $\mu$:
\bea
   A^{(1)t} & = & \frac{\alpha_s(\mu)}{4\pi} \left\{ 
   C_7^{(1)}(\mu) - \frac{16}{3} \, C_7^{(0)eff}(\mu) 
   \right. \nonumber \\
   & & \left. 
     \qquad\qquad + \sum_i^8 C_i^{(0)eff}(\mu) \, \left[
       \gamma_{i7}^{(0)} \, \ln \frac{m_b}{\mu} + r_i(z) \right] 
     \right\} \; , \\
   A^{u} & = & \frac{\alpha_s(\mu)}{4\pi} \,
     C_2^{(0)}(\mu) \, \left[ r_2(z) - r_2(0) \right] ~,
\eea
where $r_i$'s are complex numbers. Expressions for the various
quantities appearing in the above equations can be found in
Refs.~\cite{ghw,cmm}. We stress that the gluon bremsstrahlung parts
have been dropped in calculating $\Gamma(B \to \rho \gamma)$,
except those needed to cancel the divergence in the $O(\alpha_s)$
virtual corrections in the decay $b \to d \gamma$.  Note
that, in the above rate, all terms higher than $O(\alpha_s)$ have to
be dropped for theoretical consistency. The expression for $\Gamma(B^0
\to \rho^0 \gamma)$ can be obtained by obvious replacements, except
that $L_R^{u}(B^0) \ll L_R^{u}(B^\pm)$.

Using the above expression, the NLO isospin-violating ratio $\Delta$
is found to be:
\bea
  \Delta_{\rm NLO} & = & \Delta_{\rm LO} 
  \nonumber \\
  & & 
  \hskip-5truemm
- \frac{2 \epsilon_A}{\c_7^{(0)eff}} \left[
    F_1 A_R^{(1)t} - \left( F_2^2 - F_1^2 \right) A_R^u 
    + \epsilon_A \left( F_1^2 + F_2^2 \right) \left(A_R^{(1)t} + F_1 A_R^u \right)
    \right] \; .
\label{eq:deltanll}
\eea
where $\Delta_{\rm LO}$ is given in Eq.~(\ref{eq:deltall}).

\begin{figure}[b!]
     \begin{minipage}[b]{0.5\textwidth}
       \centering 
       \includegraphics[scale=0.4]{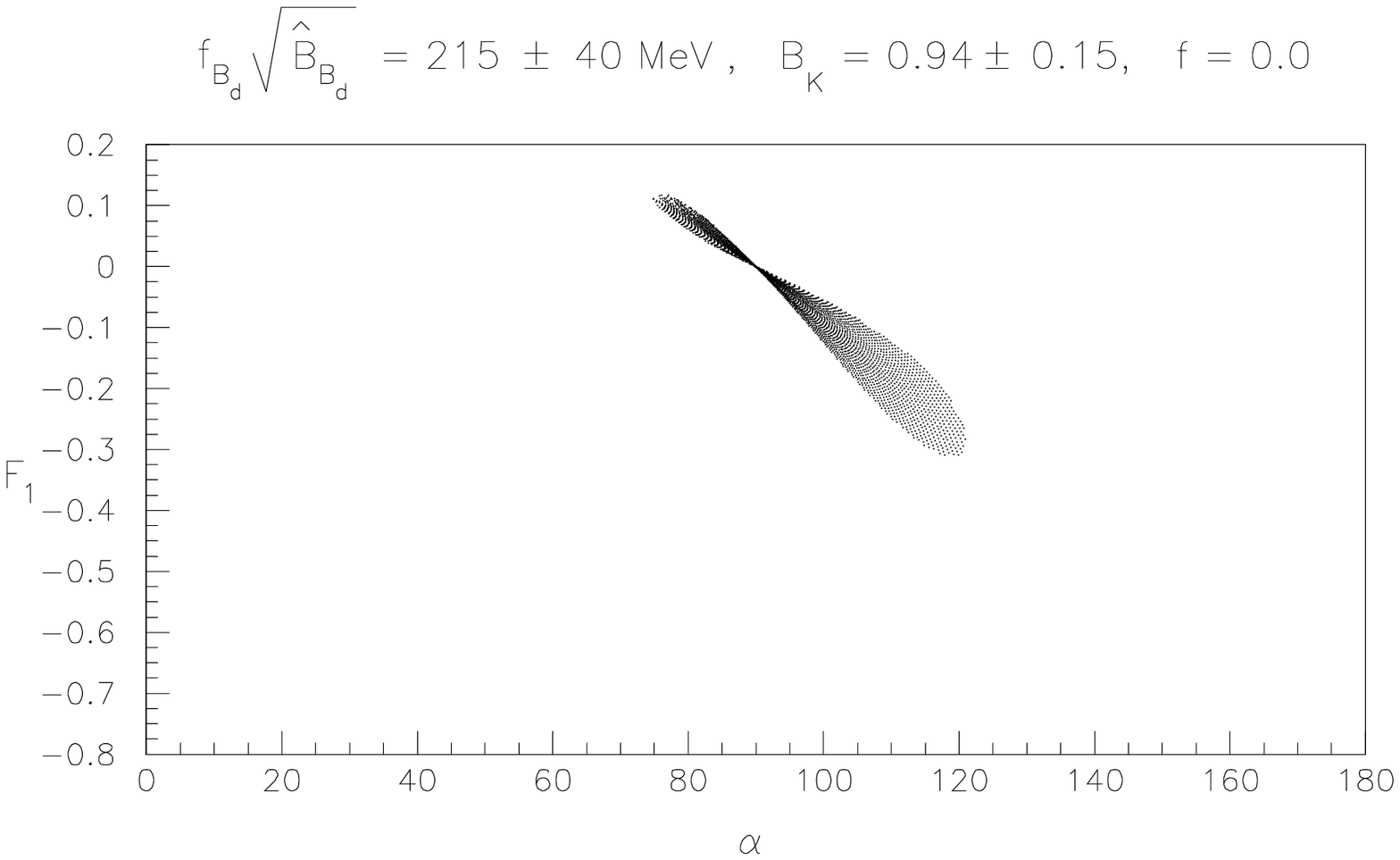} \\
        \includegraphics[scale=0.4]{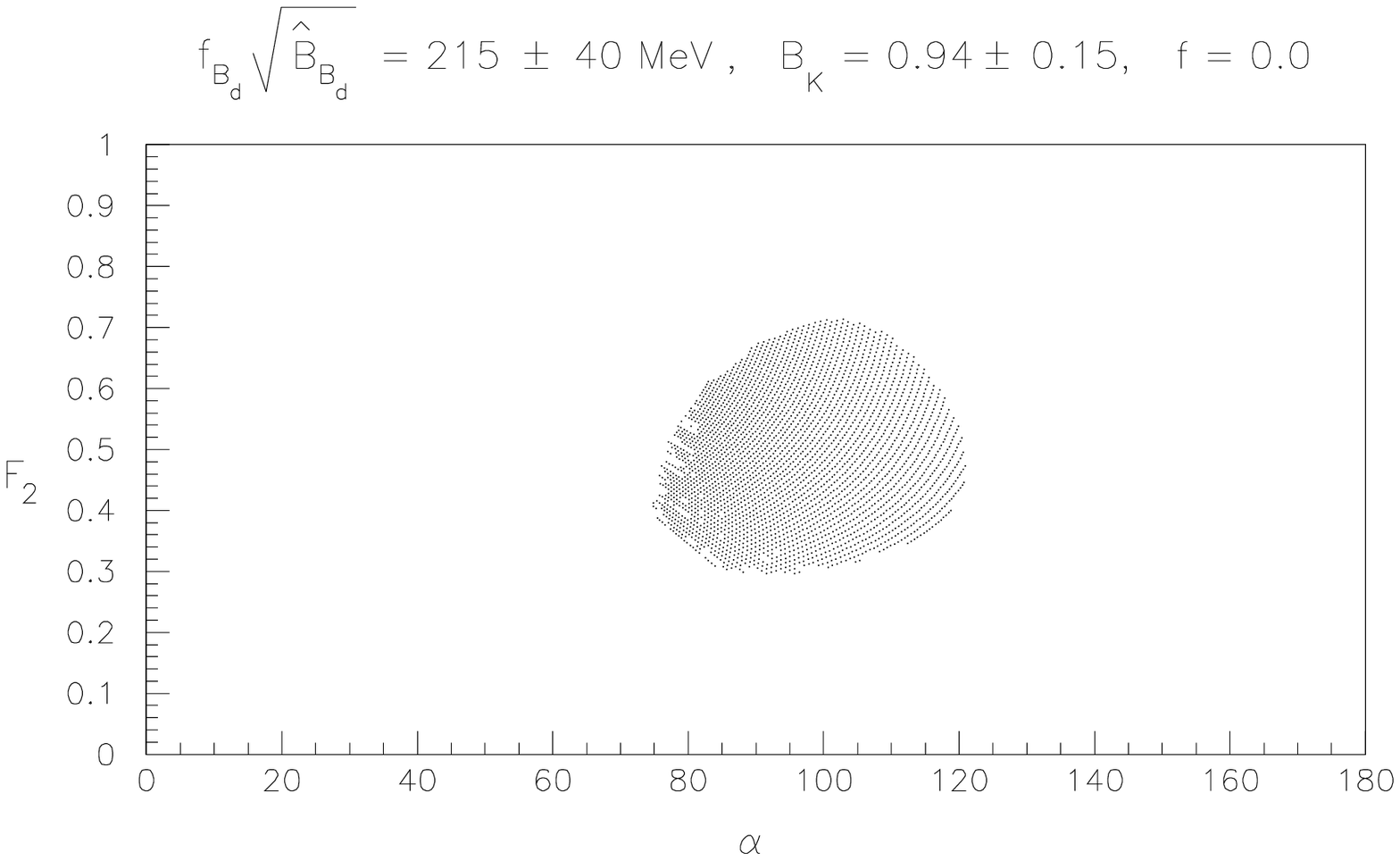}
     \end{minipage} 
     \begin{minipage}[b]{0.5\textwidth}
       \centering 
       \includegraphics[scale=0.4]{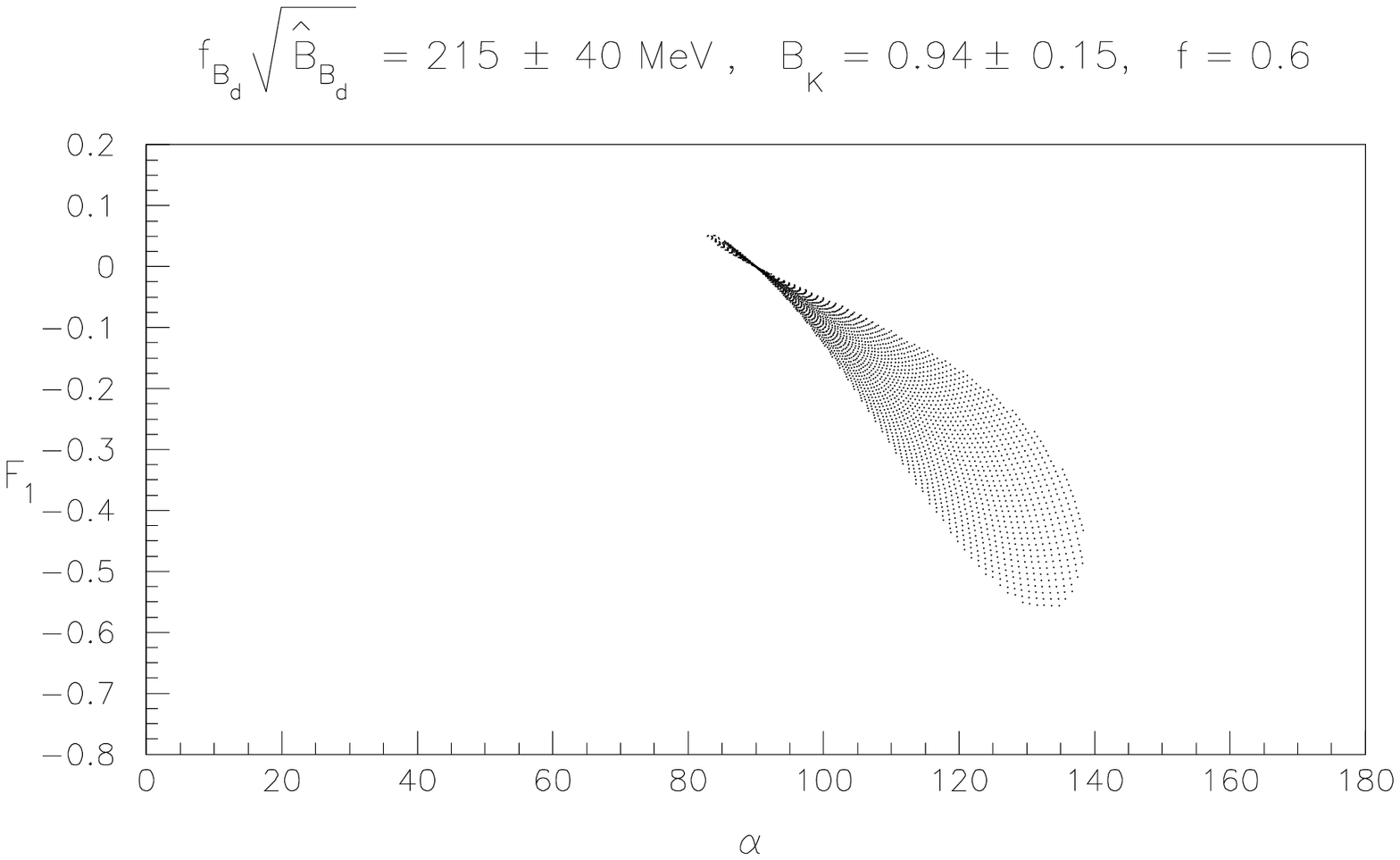}\\
        \includegraphics[scale=0.4]{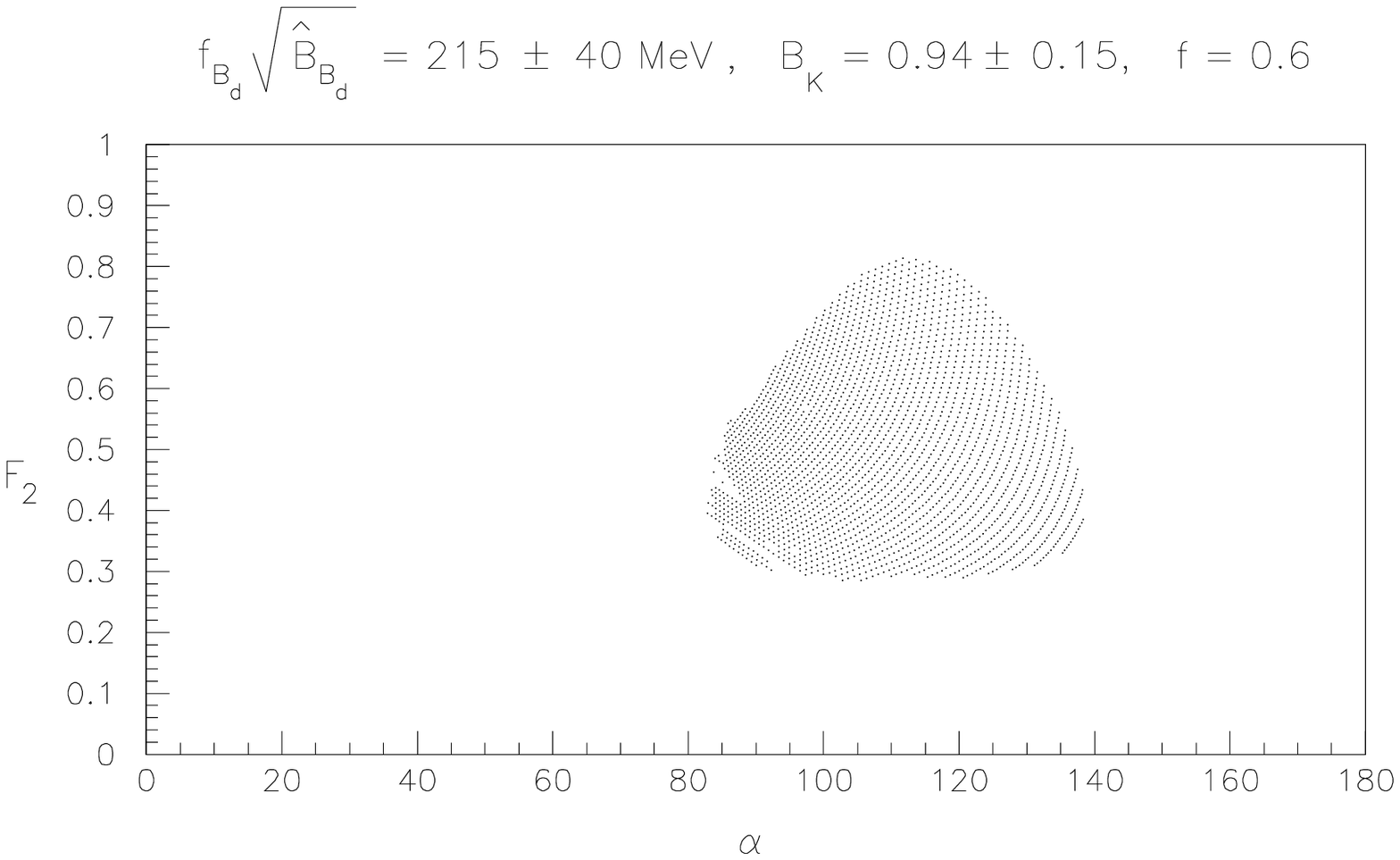}
     \end{minipage}
    \caption{Upper-left: SM $F_1$--$\alpha$ correlation. 
      Upper-right: MSSM $F_1$--$\alpha$ correlation. 
      Lower-left: SM $F_2$--$\alpha$ correlation. 
      Lower-right: MSSM $F_2$--$\alpha$ correlation. The parameters used
in calculating the correlations are indicated on top of the figures.}
    \label{F2alpha}
\end{figure}

The values for the various input quantities used in the numerical
calculations of $\Delta_{\rm LO}$ and $\Delta_{\rm NLO}$ in the SM are
as follows: $\c_7^{(0) eff}(m_b) =-0.318$, $A_R^{(1) t} =-0.022$,
$A_R^{u}=+0.049$, and $\epsilon_A = -0.3$. The remaining ingredient is
a determination of the allowed ranges for the functions $F_1$ and
$F_2$. Taking into account the present experimental and theoretical
constraints on the parameters of the CKM matrix, the profile of the
unitarity triangle in the SM was presented by two of us in
Ref.~\cite{al99}. In Fig.~\ref{F2alpha} we show the allowed
$F_1$--$\alpha$ and $F_2$--$\alpha$ correlations at 95\% C.L. In these
figures,
the SM plots are found on the left-hand side, and are labelled by
$f=0.0$. The ranges of the hadronic parameters
$f_{B_d}\sqrt{\hat{B}_{B_d}}$ and $B_K$ used in these fits are
indicated on top of the figures. (For definitions and further
discussions, see Ref.~\cite{al99}.)  Note that the CP phase $\alpha$
is constrained to lie in the range $75^\circ \le \alpha \le 121^\circ$
at 95\% C.L. \cite{al99}.

\begin{figure}[t!]
      \centering 
     \begin{minipage}[c]{0.4\textwidth}
       \centering 
       \includegraphics[scale=0.4]{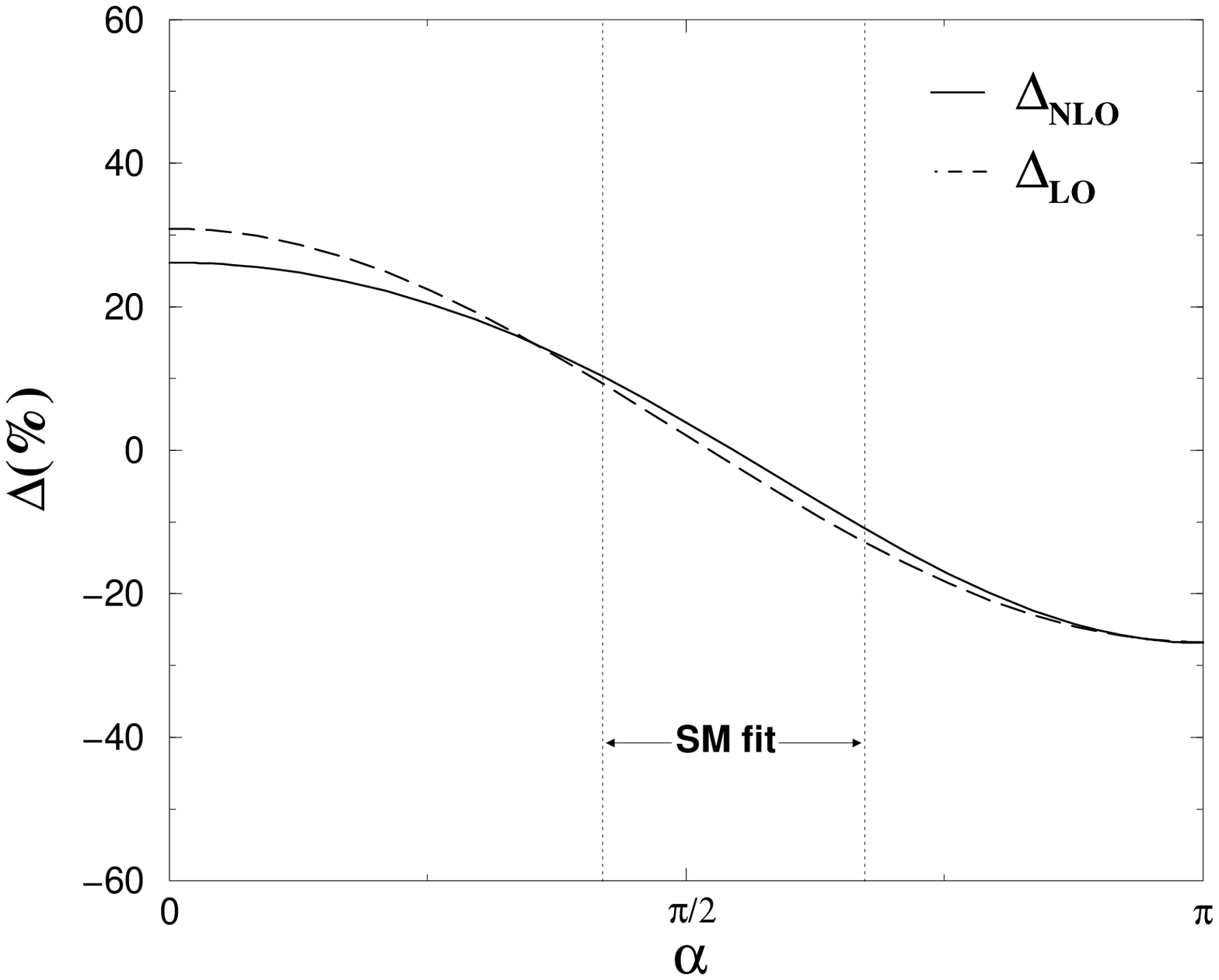}\\
     \end{minipage}
    \caption{The isospin-violating ratio $\Delta$ at LO and NLO
      in the SM. We have set $\epsilon_A=-0.3$. The curves correspond
      to the central value of the CKM fits $\vert V_{ub} / V_{td}
      \vert=0.48$ \protect\cite{al99}.}
    \label{fig:delta}
\end{figure}

With this information, we can now calculate the ratios $\Delta_{\rm
  LO}$ and $\Delta_{\rm NLO}$ in the SM. In Fig.~\ref{fig:delta} the
results are shown for these quantities as a function of the angle
$\alpha$. In these figures we have assumed that $\vert V_{ub} / V_{td}
\vert=0.48$ (its central value \cite{al99}). However, for a given
value of $\alpha$, $\Delta_{\rm LO}$ and $\Delta_{\rm NLO}$ may in
fact take a range of values. This residual CKM-related range is given
essentially by the $F_1$--$\alpha$ correlation presented in the
upper-left plot in Fig.~\ref{F2alpha}. Note that the isospin-violating
ratio is very stable against NLO corrections in the SM. This
observation, together with the discussions earlier about the
determination of $\epsilon_A$, makes $\Delta$ suitable for precision tests
of the SM. In particular, its measurement will determine $\alpha$ in the
SM.

We now turn to the direct CP asymmetry \cite{kn}. As noted earlier,
the strong interaction phase $\phi_A$ of Eq.~(\ref{phiAdef})
disappears in the chiral limit \cite{gp00}, which implies that, to
lowest order, there is no CP-violation in the decay rates for $B \to
\rho\gamma$.  Therefore, the strong phases in the exclusive decays
$B^\pm \to \rho^\pm \gamma$ and $B^0(\overline{B^0}) \to \rho^0
\gamma$, which are necessary for inducing direct CP-violation, must be
generated by higher-order perturbative QCD corrections. Concentrating
on the charged $B$ decays, we define the CP asymmetry as
\beq
        \acp \equiv
        \frac{\b(B^- \rightarrow \rho^- \gamma)
          - \b(B^+ \rightarrow \rho^+ \gamma)}{
          \b(B^- \rightarrow \rho^- \gamma)
          + \b(B^+ \rightarrow \rho^+ \gamma)} \; .
\eeq
Since, in the heavy quark limit, there are no non-factorizing strong
phases in the $W$-annihilation part of the $B \to \rho \gamma$
amplitudes \cite{gp00}, the strong phases are generated {\it entirely}
by the Bander-Silverman-Soni mechanism \cite{bss}, which involves the
interference of the penguin operator ${\cal O}_7$ and the four-quark
operator ${\cal O}_2$ \cite{soares,gsw}. This mechanism has been
employed by Greub, Simma and Wyler to calculate $\acp$, using a wave
function model \cite{gsw} for the mesons. Since we are working to
leading twist, we shall ignore the effects involving virtual
corrections off the spectator quarks, arguing that they are suppressed
by powers of $1/m_b$. In that case, the CP asymmetry is determined by
perturbation theory up to a non-perturbative quantity which can be
determined from the ratio $\Delta$. The expression for $\acp$ is given
by
\bea
  \acp & = & -\frac{2 F_2}{\c_7^{(0)eff} (1+\Delta_{\rm LO})} 
  \left[ A_I^u  - \epsilon_A A_I^{(1)t}  \right] ~.
\label{eq:acp}
\eea
The quantities $A_I^{(1)t}$ and $A_I^u$ take the values
$A_{I}^{(1)t}=-0.016$
in the SM and $A_{I}^{u}=+0.046$ in both the SM and MSSM.

Note that $\Delta$ and $\acp$ are complementary measurements. Dropping
the small $O(\epsilon_A^2)$ terms in Eq.~(\ref{eq:deltanll}), we see
that $\Delta$ is essentially proportional to $F_1$, while $\acp$ is
proportional to $F_2$. Thus, for $\alpha \simeq \pi/2$, $\Delta$ is
very small, while $\acp$ takes its maximal value. Conversely, if the
value of $\alpha$ is far from $\pi/2$, the CP asymmetry decreases,
while $\Delta$ becomes measurable.

We are now ready to examine the supersymmetric contributions to
$\Delta$ and $\acp$. To begin with, we note that the NLO corrections
to the decays $B \to X_s \gamma$ have been calculated in only one
particular realization of the MSSM, the so-called {\it minimal flavor
  violation} scenario \cite{cdgg}. While this calculation considers an
important parameter space in the MSSM, it nevertheless neglects other
contributions, such as those from gluinos, which are important in
other regions of parameter space \cite{bghw}. In the small-$\tan
\beta$ domain, where the neglected contributions are small, we have
numerically calculated the NLO quantities and found that the NLO
correction to $\Delta$ in the MSSM with minimal flavor violation is
very similar to that in the SM, and hence unimportant. The complete
NLO corrections for the large-$\tan \beta$ case, including gluino
contributions, are not yet available. Hence, in comparing the SM
profile with that of the MSSM, we shall restrict ourselves to
$\Delta_{\rm LO}$.

Supersymmetry can affect $\Delta$ and $\acp$ in two distinct ways.
First, the allowed values of the functions $F_1$ and $F_2$ are
different in the MSSM. We recall that the supersymmetric contributions
to the mass differences $M_{12}(B)$ and $M_{12}(K)$ can be written as
follows (for details and references, see Ref.~\cite{al99}):
\bea
\delmd &=& \delmd (SM) [ 1 +
f_d(m_{\chi_2^\pm},m_{\tilde{t}_R},
m_{H^\pm}, \tan \beta) ], \nonumber \\
\delms &=& \delms (SM) [ 1 +
f_s(m_{\chi_2^\pm},m_{\tilde{t}_R},
m_{H^\pm}, \tan \beta) ], \nonumber \\
\abseps &=& \frac{G_F^2f_K^2M_KM_W^2}{6\sqrt{2}\pi^2\Delta M_K}
\hat{B}_K\left(A^2\lambda^6\bar{\eta}\right)
\left(y_c\left\{\hat{\eta}_{ct}f_3(y_c,y_t)-\hat{\eta}_{cc}\right\}
\right. \nonumber \\
&~& \left. +
~\hat{\eta}_{tt}y_tf_2(y_t)[1 + f_\epsilon
(m_{\chi_2^\pm},m_{\tilde{t}_2}, m_{H^\pm},
\tan \beta)] A^2\lambda^4(1-\bar{\rho})\right).
\label{susyformel}
\eea
To an excellent approximation, one has $f_d = f_s = f_\epsilon \equiv
f$. The quantity $f$ is a function of the masses of the (lighter) 
right-handed top squark
($m_{\tilde{t}_R}$), chargino ($m_{\tilde{\chi}^\pm_2}$) and the
charged Higgs ($m_{H^\pm}$), as well as of $\tan\beta$. The maximum
allowed value of $f$ depends on the model. Typical values are: minimal
supergravity ($f=0.2$), non-minimal supergravity ($f=0.4$)
\cite{goto99}, and MSSM with constraints from electric dipole moments
(EDM's) ($f=0.6$) \cite{koetal}. The plots in the upper right-hand and
lower right-hand corners of Fig.~\ref{F2alpha} show the allowed
$F_1$--$\alpha$ and $F_2$--$\alpha$ correlations, respectively, for
the MSSM with $f=0.6$. We see that these correlations can be
measurably different from the SM. In particular, much larger values of
$\alpha$ are allowed compared to the SM. Thus, for $f=0.6$, the fits
yield $86^\circ \le \alpha \le 141^\circ$ at 95\% C.L.

The second way in which supersymmetry affects $\Delta$ and $\acp$ is
via the Wilson coefficients. In contrasting the SM and MSSM profiles,
we assume, as per the usual expectations, that the coefficients of the
tree amplitudes, $\c_1^{(0)}$ and $\c_2^{(0)}$, are the same in these
models, but that $\c_7^{(0)eff} (\mu)$ may differ. This latter
coefficient is constrained by the measured branching ratio of the
decay $B \to X_s \gamma$, yielding a bound $2.0 \times 10^{-4} \leq
{\cal B}(B \to X_s \gamma) \leq 4.5 \times 10^{-4}$ at 95\% C.L.
\cite{cleobsg}. The resulting constraints on the magnitude and phase
of the ratio $\c_7^{(0)eff}/\c_7^{(0)eff\rm (SM)}$ in the context of
the MSSM being considered can be summarized as follows. In the absence
of the constraints on the EDM's of the neutron and electron, the real
and imaginary parts of $\c_7^{(0)eff}/\c_7^{(0)eff\rm(SM)}$ can vary
substantially.  But if one takes into account the EDM constraints, the
imaginary part of $\c_7^{(0)eff}/\c_7^{(0)eff\rm (SM)}$ is highly
suppressed. However, depending on the value of $\tan \beta$, both
positive- and negative-valued solutions of $\c_7$ are allowed. For
example, a recent analysis of $\c_7^{(0)eff}/\c_7^{(0)eff\rm (SM)}$ in
the minimal supergravity model yields values in the ranges $0.7 \leq
{\rm Re} [\c_7^{(0)eff}/\c_7^{(0)eff\rm{(SM)}}] \leq 1.2$ for small
$\tan \beta$ (say $\tan \beta \leq 10$), but for larger values of
$\tan \beta$, negative values of this ratio are admissible. Thus, for
$\tan \beta =30$, a range $-1.5 \leq
\c_7^{(0)eff}/\c_7^{(0)eff\rm(SM)} \leq -0.8$ is allowed by present
data \cite{goto99}.

As for $\acp(B^\pm \to \rho^\pm \gamma)$, the dominant contribution,
proportional to $A_I^u$, is identical for the SM and MSSM. However,
the value of the quantity $A_I^{(1)t}$ in the MSSM depends on the
region of parameter space considered. For small $\tan \beta$,
$A_I^{(1)t}$ has almost the same value as in the SM, while for large
$\tan \beta$, it may appreciably differ from the SM value. However,
since its contribution to $\acp(B^\pm \to \rho^\pm \gamma)$ is
suppressed due to the $\epsilon_A$ factor, its precise value is not so
important numerically. In calculating $\acp$ in the MSSM, we have set
$A_I^{(1)t}=0$.

\begin{figure}[t!]
      \centering 
     \begin{minipage}[c]{0.4\textwidth}
       \centering 
       \includegraphics[scale=0.4]{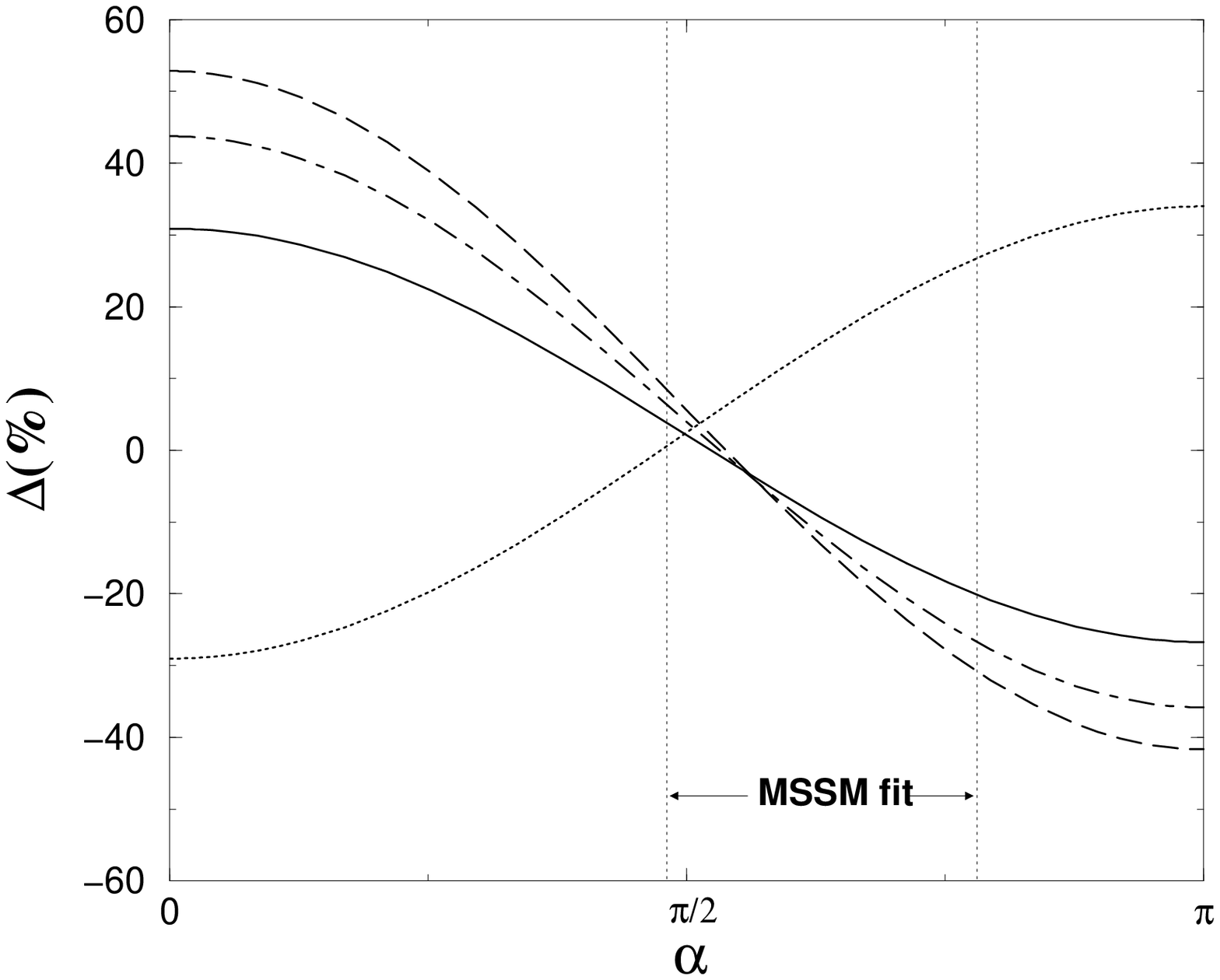}
     \end{minipage} 
     \hspace*{1cm}
     \begin{minipage}[c]{0.4\textwidth}
       \centering 
       \includegraphics[scale=0.4]{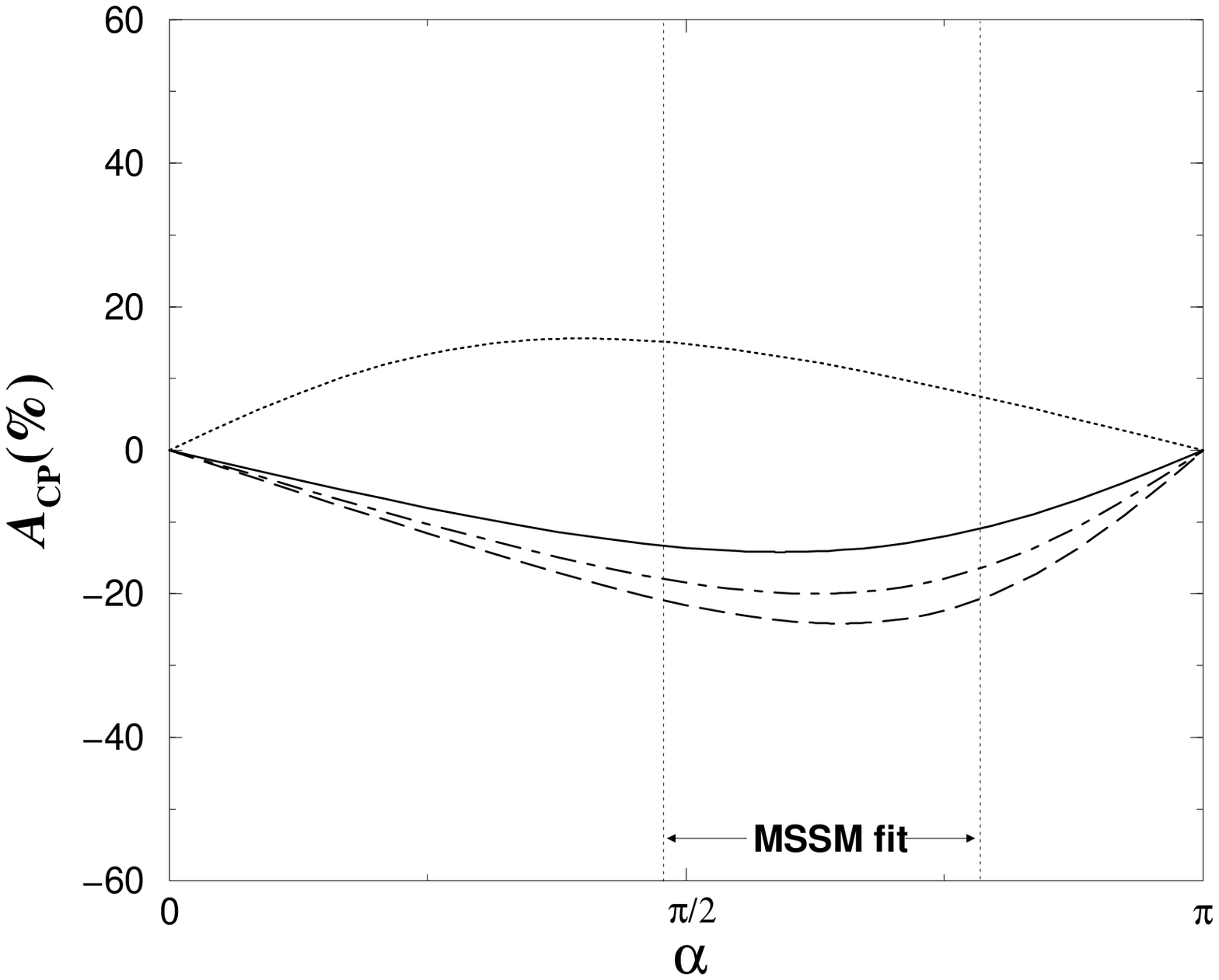}
     \end{minipage}
    \caption{The isospin-violating ratio $\Delta$ in LO (left) and 
      $\acp(B^\pm \to \rho^\pm \gamma)$ (right) with $\epsilon_A=-0.3$
      in the SM (solid
      line), and in the MSSM with $(\tan \beta,
      {\c_7^{(0)eff}}/{\c_7^{(0)eff\rm (SM)}}) = (3,0.95)$ (dot-dashed
      line), $(\tan \beta, {\c_7^{(0)eff}}/{\c_7^{(0)eff\rm (SM)}}) =
      (30,0.8)$ (dashed line) and $(\tan \beta,
      {\c_7^{(0)eff}}/{\c_7^{(0)eff\rm (SM)}}) = (30,-1.2)$ (dotted
      line). The SM and MSSM curves
      correspond respectively to $\vert V_{ub} / V_{td} \vert=0.48$
      and $\vert V_{ub} / V_{td} \vert=0.63$, which are the central
      values in the CKM fits \protect\cite{al99}. The allowed ranges
      for $\alpha$ in the MSSM from these fits for $f=0.6$ are also
      indicated.  }
\label{deltasusy}
\end{figure}

In Fig.~\ref{deltasusy} we contrast the expectations for $\Delta$ in
the SM and in two variants of the MSSM, characterized by small and
large values of $\tan \beta$.  The residual CKM-related range in the
allowed values of $\Delta$ is again given essentially by the
$F_1$--$\alpha$ correlation presented earlier in the upper two plots
of Fig.~\ref{F2alpha} for the SM ($f=0.0$) and MSSM ($f=0.6$).  Note
that if the large $\tan \beta$ MSSM solution is realized in nature,
the measured value of $\Delta^{\pm 0}$ can be markedly different than
in the SM. More importantly, since ${\c_7^{(0)eff}}/{\c_7^{(0)eff\rm
    (SM)}}$ can be negative, the sign of $\Delta^{\pm 0}$ may change
over a large region of the allowed CKM parameter space. This would be
a striking signature of new physics, and would strongly suggest the
presence of supersymmetry.

Similar effects appear in the CP asymmetry $\acp$. Since, in the
models being considered here, there are no other phases at this order,
Eq.~(\ref{eq:acp}) holds in the MSSM, with the proviso that
numerically $\c_7^{(0)eff}$ now depends on the parameters of the MSSM.
In particular, for large $\tan \beta$, the CP asymmetry can be
significantly larger than the one in the SM. Again, since
$\c_7^{(0)eff}(MSSM) \simeq - \c_7^{(0)eff}(SM)$ is allowed, the CP
asymmetry $\acp(B^\pm \to \rho^\pm \gamma)$ reverses sign in this case
(as does the asymmetry in the inclusive decays $\acp(B^\pm \to X_d^\pm
\gamma)$). This is illustrated in Fig.~\ref{deltasusy}. Once again,
this would be a clear signal of supersymmetry in the large $\tan
\beta$ domain.

In summary, we have examined the effects of supersymmetry on two
observables of $B\to \rho\gamma$ decays: the isospin-violating
quantity $\Delta$ and the direct CP asymmetry $\acp$. We find that, in
MSSM models, the predictions for these quantities can be substantially
modified. In particular, over a large region of parameter space, the
signs of $\Delta$ and $\acp$ may be flipped, which would be a clear
signal of new physics, and would point directly to the presence of
supersymmetry.

\medskip
We thank Christoph Greub, Toru Goto, Gudrun Hiller, Klaus Honscheid,
David Jaffe, Alexey Petrov and Dan Pirjol for helpful discussions and
communication. The work of D.L. was financially supported by NSERC of
Canada.

\end{document}